\documentclass[final,5pt,twocolumn]{elsarticle}

\usepackage{amssymb}

\journal{Physica C}

\newcommand{\Vect}[1]{\mbox{\boldmath$#1$}}

\begin{document}

\begin{frontmatter}



\title{Non-equilibrium superconductivity in a correlated electron system
studied with the Keldysh+FLEX approach
}


\author{Takashi Oka, Hideo Aoki}

\address{Department of Physics, the University of Tokyo, Hongo, Tokyo 113-0033, Japan}

\begin{abstract}
Non-equilibrium phase transitions are studied theoretically 
for the two-dimensional Hubbard model subject to 
bias voltages from the electrodes coupled to the system. 
By combining the fluctuation exchange approximation 
with the Keldysh method for non-equilibrium, we
have studied the properties of the non-equilibrium Fermi
liquid phase and determined the phase diagram
with transition to non-equilibrium magnetic and 
superconducting phases.
\end{abstract}

\begin{keyword}
Nonequilibrium superconductivity \sep Hubbard model
 \sep 74.40.+k \sep 05.30.-d \sep 71.10.-w

\end{keyword}

\end{frontmatter}


\section{Nonequilibrium Superconductivity and Electron Correlation}
\label{section1}

\begin{figure}[tbh]
\centering 
\includegraphics[width=5cm]{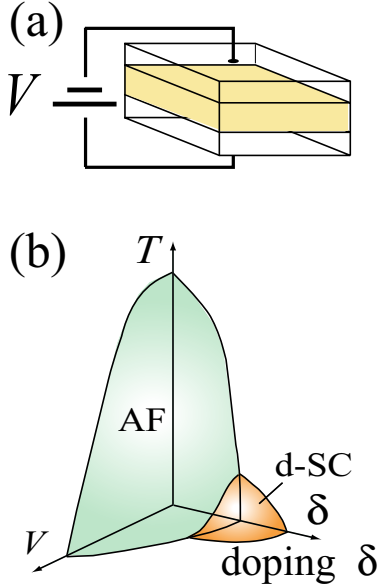}
\caption{(a) Two dimensional correlated system (shaded) 
in a finite bias $V$.  
(b) Obtained non-equilibrium phase diagram against the bias ($V$), 
temperature ($T$), and the doping level ($\delta$) 
with antiferromagnetic (AF) and d-wave superconducting (SC) phases.
}
\label{fig}
\end{figure}

Due to the decades of research for 
the HTC cuprates, organic superconductors and other 
classes of materials, the understanding of correlated 
electron systems is becoming matured.  
One novel avenue to pursue is 
alternative methods for doping other than the conventional chemical 
doping.  One way is to inject carriers from 
interfaces or electrodes 
which has been recently realized
in oxides heterostructures\cite{Exps}. 
If we consider the case where two electrodes 
are attached, then by adjusting the bias voltage
across the electrodes, we can introduce 
carriers whose distribution may be different 
from equilibrium.
The effect of such non-equilibrium 
carriers in correlated electron systems 
has been theoretically proposed recently by the
present authors \cite{OkaSC09}.  
The main interest is 
whether superconductivity can occur in non-equilibrium 
in an open, two-dimensional correlated electron system coupled to 
electrodes (Fig.~\ref{fig}~(a)). We have adopted 
the Hubbard model as a standard model for 
correlated electron systems.  The non-equilibrium 
situation is treated with the Keldysh formalism, while 
the many-body effect with the FLEX approach\cite{OkaSC09}.  
We also note that non-equilibrium methods for
doping cuprates have been experimentally 
realized in ref. \cite{Exps2}
where photoinduced superconductivity in HTC materials
was observed. Our approach is expected to shed light
to these systems as well.

\section{Nonequilibrium Phase Diagram and Superconducting Order}
\label{section2}

\begin{figure}[tbh]
\centering 
\includegraphics[width=7.5cm]{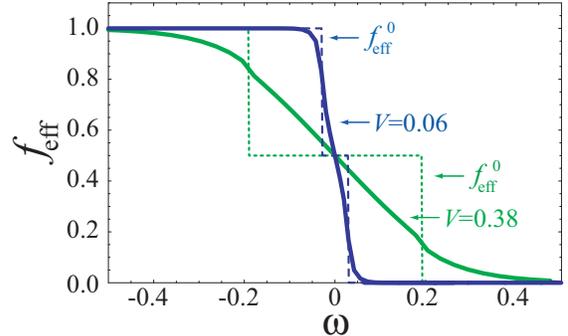}
\caption{Nonequilibrium distribution function
for two values of the bias $V$ at half filling ($\delta=0$).
Dashed lines 
are the noninteracting distribution function $f_{\rm eff}^0$. 
}
\label{fig15}
\end{figure}

We consider a thin layer of strongly correlated material 
described by the two-dimensional Hubbard model which is 
coupled to electrodes.  
Here we assume that the electrodes are attached in the
vertical direction (Fig.~\ref{fig}~(a)) and double step Fermi
distribution is realized in the correlated system (Fig.~\ref{fig15}). 

In non-equilibrium systems, an important effect
of the electron-electron interaction is the smearing
of the electron distribution function. 
As can be seen in Fig.~\ref{fig15},
the double step Fermi distribution
$f_{\rm eff}^0$ realized by the electrode and 
bias $V$ becomes smeared into a smooth 
non-equilibrium distribution function $f_{\rm eff}$.
In this {\it non-equilibrium Fermi liquid} phase 
the bias voltage $V$ effectively 
plays a similar role as the temperature
since the smearing of the distribution becomes
stronger as one increase $V$.

\begin{figure}[tbh]
\centering 
\includegraphics[width=8.5cm]{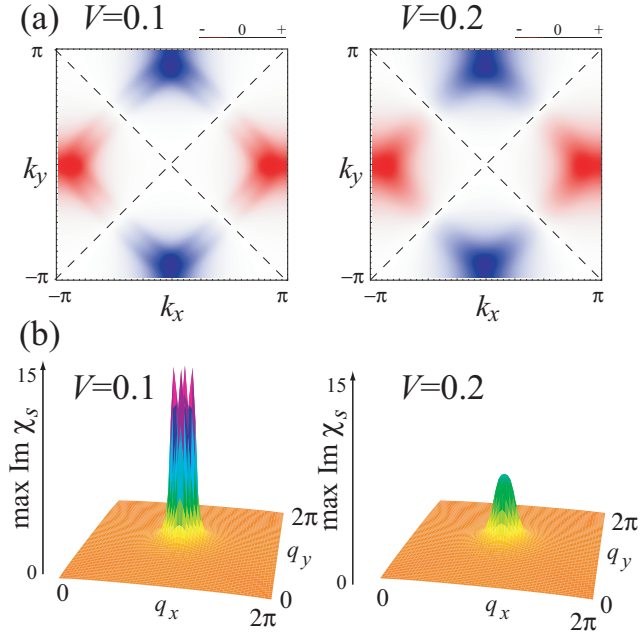}
\caption{
(a) Superconducting gap function
$\mbox{Re}\,\phi(\Vect{k},\omega= 0)$ in non-equilibrium.
The bias is $V=0.1,\;0.2$.  Dashed lines represent nodes.  
 (b) Spin susceptibility $\chi(\Vect{q},\omega=0)$ 
for $V=0.1,\;0.2$.
For $\delta=0.14$, 
$U=4.5$, and $\mu=-0.35$.
}
\label{fig2}
\end{figure}

The non-equilibrium phase diagram obtained within
the Keldysh + FLEX method
is schematically shown in Fig. \ref{fig} (b). 
Away from half-filling, for which FLEX is expected to be applicable, 
the superconducting gap function 
with  $d$-wave symmetry dominates even 
in finite bias voltages.
However, as one increases the bias,
we can see that both the AF and 
SC phases become suppressed.  
Indeed, this can be seen in Fig.~\ref{fig2}~(b).
First at bias $V=0.1$, which is near the 
non-equilibrium Fermi liquid to d-wave superconductor
transition, the AF fluctuation is sharp
and has multiple peak structure which reflects the
incommensurate nesting. However, as one 
goes far away from the transition ($V=0.2$), the 
peak becomes smaller and the structure vanishes.

The suppressed orderings can be understood 
in terms of the change in the 
non-equilibrium electron distribution.
The AF fluctuation is induced by electron-electron 
scattering across the Fermi surface 
which is enhanced near half-filling due to nesting effect.
However, in the presence of finite bias $V$, (i) the 
Fermi surface becomes split into two with 
Fermi energies $E_F\pm V/2$ where $E_F$ is the
original Fermi energy, and (ii) 
smearing takes place. The first effect reduces the 
nesting while the second weakens the scattering.
The AF fluctuation is the glue for the transition
from (non-equilibrium) Fermi liquid
to d-wave superconductivity and the reduction of
the fluctuation results in the destruction 
of the superconducting phase.

In conclusion, we have studied the 
transition between the non-equilibrium Fermi liquid, 
AF magnetic order and d-wave superconductor 
which is controllable by the applied bias. 
Finally we comment on the smearing effect 
which destroys the long range order in finite bias.
In the Keldysh + FLEX method, the self-consistent 
treatment of the Green's function tends 
to overestimate the smearing effect. 
If the smearing is not as strong, there
is a possibility that a new order 
appears which does not exist in equilibrium \cite{OkaSC092}.

\end{document}